\documentclass[aps,preprint,prd,superscriptaddress,amsfont,graphicx,nofootinbib,preprintnumbers,11pt]{revtex4-2}

\usepackage{color,graphicx,epsfig}
\usepackage{ifpdf}
\usepackage{amsmath}
\usepackage{bm}
\usepackage{color}
\usepackage[english]{babel}
\usepackage{graphicx}%
\usepackage{amsfonts}%
\usepackage{amssymb}
\usepackage{braket}
\usepackage{hyperref}

\usepackage{enumitem}
\setlist[itemize]{itemsep=0pt,topsep=\parsep}
\setlist[enumerate]{itemsep=0pt,topsep=\parsep}

\usepackage{ulem}

\definecolor{nicered}{rgb}{0.7,0.1,0.1}
\definecolor{nicegreen}{rgb}{0.1,0.5,0.1}
\hypersetup{colorlinks,citecolor= nicegreen,linkcolor= nicered}

\makeatletter
\g@addto@macro\bfseries{\boldmath}
\let\Hy@backout\@gobble
\makeatother

\newenvironment{Eqnarray}{\arraycolsep 0.14em\begin{eqnarray}}{\end{eqnarray}}
\def\beqa{\begin{Eqnarray}}
\def\eeqa{\end{Eqnarray}}
\newcommand{\no}{\nonumber}

\newcommand{\beq}{\begin{equation}}
\newcommand{\eeq}{\end{equation}}
\newcommand{\bea}{\begin{eqnarray}}
\newcommand{\eea}{\end{eqnarray}}

\begin{document}

\title{CP asymmetries in charged meson decay to two pions}

\author{Yuval Grossman}
\email{yg73@cornell.edu}
\affiliation{LEPP, Department of Physics, Cornell University, Ithaca, NY 14853, USA}

\author{Zoltan Ligeti}
\email{ligeti@berkeley.edu}
\affiliation{Theory Group, Lawrence Berkeley National Laboratory, Berkeley, CA 94720, USA}
\affiliation{\mbox{Leinweber Institute for Theoretical Physics, University of California, Berkeley, CA 94720, USA}}

\author{Yosef Nir}
\email{yosef.nir@weizmann.ac.il}
\affiliation{Department of Particle Physics and Astrophysics, Weizmann Institute of Science, Rehovot 7610001, Israel}

\begin{abstract}

We study the CP asymmetries in $B^+$, $D^+$, and $K^+$ decays into $\pi^+\pi^0$. These asymmetries are commonly known to vanish in the isospin limit. We clarify what is meant by the term ``isospin limit" in this context. We provide a unified formalism to discuss these asymmetries, and develop an understanding of how various suppression factors distinguish between them qualitatively. 
We estimate the size of the asymmetries in the Standard Model:  $A_{CP}(B^+\to\pi^+\pi^0)\sim3\times10^{-3}$,  $A_{CP}(D^+\to\pi^+\pi^0)\sim10^{-5}$, and \mbox{$A_{CP}(K^+\to\pi^+\pi^0)\sim10^{-6}$}.

\end{abstract}

\maketitle

\newpage

%%%%%%%%%%%%%
%%%%%%%%%%%%%
%%%%%%%%%%%%%
\section{Introduction}
None of the three CP asymmetries,
\beq\label{eq:acpdef}
A_{CP}(P^\pm\to\pi^\pm\pi^0)\equiv\frac{{\cal B}(P^-\to\pi^-\pi^0)-{\cal B}(P^+\to\pi^+\pi^0)}{{\cal B}(P^-\to\pi^-\pi^0)+{\cal B}(P^+\to\pi^+\pi^0)},
\qquad P=B,D,K,
\eeq
has been experimentally established to be different from zero, and 
only bounds are available~\cite{ParticleDataGroup:2024cfk}. For the $B$ decay they are from measurements by BaBar \cite{BaBar:2007uoe}, Belle \cite{Belle:2012dmz} and Belle II \cite{Belle-II:2023ksq}, for the $D$ decay from measurements by LHCb \cite{LHCb:2021rou}, Belle \cite{Belle:2017tho}, CLEO \cite{CLEO:2009fiz}, and  Belle~II \cite{Belle-II:2025wsy},
while for the $K$ decay there is a single measurement from 1969 \cite{Herzo:1969tyz}.
The current experimental limits, together with the respective CP-averaged branching ratios, are presented in Table~\ref{tab:acpexp}.

How far are these measurements from the Standard Model (SM) predictions? It is common knowledge that the three asymmetries vanish in the isospin limit. Consequently, in addition to possible CKM suppression, they are expected to be suppressed by the small isospin violation of ${\cal O}(0.01)$. 

Our three main goals in this study of $A_{CP}(P^\pm\to\pi^\pm\pi^0)$ are the following:
\begin{enumerate}
    \item Clarify what is meant by ``isospin violation" for these decays.
    \item Provide a unified formalism to qualitatively discuss the CP asymmetries in $B^\pm$, $D^\pm$,  and $K^\pm$ decays into $\pi^\pm\pi^0$, and in parallel develop an understanding of how various suppression factors distinguish between them quantitatively.
    \item Calculate specific contributions to the CP asymmetries, while pointing out additional sources that are not calculable with current symmetry-related tools.    
\end{enumerate}

The term ``isospin limit" needs clarification in the case of $P\to\pi\pi$ decays. 
\begin{itemize}
    \item Under the $SU(2)_I$ isospin symmetry, the initial state is $I=1/2$, while the final state is $I=0$ or $2$. Thus, imposing isospin on the SM Lagrangian, these processes would be forbidden altogether. 
\end{itemize}
This literal interpretation is obviously not how the term ``isospin limit" is used in the relevant literature. Given that, within the SM, isospin is broken by the electroweak interactions and quark masses, and is only an approximate symmetry of the strong interactions, one finds in the literature two different definitions of what is meant by the ``isospin limit":
\begin{itemize}
\item Switching off isospin breaking in flavor conserving interactions, and considering only tree and strong penguin operator contributions to flavor changing processes.
\item Switching off isospin breaking in flavor conserving interactions.
\end{itemize}
In the first definition (and unlike in the literal interpretation), the processes are allowed. However, as is well known and reviewed below, the CP asymmetry in $P^\pm\to\pi^\pm\pi^0$ vanishes. Although we prefer to use the second definition for the term ``isospin limit", in Section~\ref{sec:truncate2} we do truncate the effective Hamiltonian in accordance with the first definition. 
However, it is important to understand that the approximation of omitting the electroweak penguin (EWP) operators has nothing to do with imposing isospin as a symmetry of the Lagrangian.

The plan of this paper is as follows. In Section~\ref{sec:truncate}, we explain why the CP asymmetry vanishes when the effective Hamiltonian ${\cal H}_{\rm eff}$ is truncated by tree and strong penguin operators, and isospin breaking in QCD is switched off. In Section~\ref{sec:EWP} we study the contribution of the EWP operators to the asymmetry. In Section~\ref{sec:isospinbreaking} we study the contribution of isospin breaking in QCD and QED to the asymmetry. Sections~\ref{sec:B}, \ref{sec:D} and \ref{sec:K} apply the universal formalism of the previous sections to the specific cases of $B^+$, $D^+$ and $K^+$, respectively. In Section~\ref{sec:conclusions} we present our conclusions. 

%%%%%%%%%%%%%
\begin{table}[t]
\begin{tabular}{|c|c|c|} \hline\hline
\rule{0pt}{1.2em}%
$P$ & ${\cal B}$ & $A_{CP}$  \cr
\hline 
$B$ & $(5.33\pm0.27)\times10^{-6}$ & $-0.012\pm0.031$ \\
$D$ & $(1.247\pm0.033)\times10^{-3}$ & $+0.011\pm0.007$ \\
$K$ & $(2.067\pm0.008)\times10^{-1}$ & $-0.004\pm0.006$ \\
\hline\hline
\end{tabular}
\caption{Experimental values of CP-averaged branching fractions and CP asymmetries for the charged meson decays $P^+\to\pi^+\pi^0$ for $P=B,D,K$ \cite{ParticleDataGroup:2024cfk,HeavyFlavorAveragingGroupHFLAV:2024ctg}. (For $A_{CP}(D^\pm\to\pi^\pm\pi^0)$, the PDG value is averaged with the recent Belle~II result~\cite{Belle-II:2025wsy}.)
Note that for $P=D,K$, Ref.~\cite{ParticleDataGroup:2024cfk} uses a sign convention opposite to Eq.~(\ref{eq:acpdef}).} 
	\label{tab:acpexp}
\end{table}   
%%%%%%%%%%%%%%%%%%

%%%%%%%%%%%
%%%%%%%%%%%
%%%%%%%%%%%
\section{General Formalism}
\label{sec:truncate}

\subsection{Isospin decomposition}
The rotation from the physical basis to the isospin basis is given by
\beqa
\langle\pi^+\pi^-|&=&\sqrt{\frac13}\langle2,0|+\sqrt{\frac23}\langle0,0|,\no\\
\langle\pi^0\pi^0|&=&\sqrt{\frac23}\langle2,0|-\sqrt{\frac13}\langle0,0|,\no\\
\langle\pi^+\pi^0|&=&\langle2,1|.
\eeqa
The decomposition of the corresponding decay amplitudes is given by~\cite{Gronau:1990ka,Botella:2006zi}
\begin{align}
A_{+-} &= -\sqrt{\frac{1}{3}}\,A_{1/2}
+ \sqrt{\frac{1}{6}}\,A_{3/2}
- \sqrt{\frac{1}{6}}\,A_{5/2}, \\
A_{00} &= \sqrt{\frac{1}{6}}\,A_{1/2}
+ \sqrt{\frac{1}{3}}\,A_{3/2}
- \sqrt{\frac{1}{3}}\,A_{5/2}, \\
A_{+0} &= \frac{\sqrt{3}}{2}\,A_{3/2}
+ \sqrt{\frac{1}{3}}\,A_{5/2},
\end{align}
where the sub-indices on the RHS correspond to the $\Delta I$ generated by operators in the effective Hamiltonian ${\cal H}_{\rm eff}$.

Using CKM unitarity, each decay amplitude $A_X$ ($X = +-, 00, +0$ for the mass basis, $X=1/2, 3/2, 5/2$ for the isospin basis) can be written as a sum of two contributions proportional to two different CKM factors~\cite{Grossman:1997gd}, which we denote by $\lambda_T$ and $\lambda_P$:
\begin{equation}
A_X = A_X^T \lambda_T + A_X^P \lambda_P \,, \qquad
\overline{A}_X = A_X^T \lambda_T^* + A_X^P \lambda_P^* \,,
\end{equation}
where $A_X^T$ and $A_X^P$ are complex hadronic amplitudes whose phases correspond to strong phases. It is convenient to choose $\lambda_T$ such that tree operators contribute only to $\lambda_T$, while penguin operators contribute to both $\lambda_T$ and $\lambda_P$. Explicitly, we choose
\begin{equation}
\lambda_T =
\begin{cases}
V_{ub}^* V_{ud} & \text{for } B, \\
V_{cd}^* V_{ud} & \text{for } D, \\
V_{us}^* V_{ud} & \text{for } K,
\end{cases}
\qquad
\lambda_P =
\begin{cases}
V_{tb}^* V_{td} & \text{for } B, \\
V_{cb}^* V_{ub} & \text{for } D, \\
V_{ts}^* V_{td} & \text{for } K.
\end{cases}
\end{equation}
Interference between the two gives the weak phase that is necessary for a direct CP asymmetry,
\beq
A_{CP}(P\to\pi\pi)\propto{\cal I}m(\lambda_{P}/\lambda_{T}).
\eeq
%

%%%%%%%%%%%
%%%%%%%%%%%
%%%%%%%%%%%

\subsection{The operator basis}

We work within the general framework of hadronic decays. The decay amplitude is written as
\begin{equation}
A = \langle out |{\cal H}_{\text{eff}} |in \rangle.
\end{equation}
The effective Hamiltonian is given by
\begin{equation}
{\cal H}_{\text{eff}}
= \frac{G_F}{\sqrt{2}}\, \lambda_X
\left(\sum_{i} c_i\, Q_i \right),
\end{equation}
where $\lambda_X$ is a linear combination of the relevant CKM factor discussed above, $c_i$ are the Wilson coefficients, and $Q_i$ are local operators.
The operators are defined at a scale of order the mass of the decaying meson when it is heavy, or at some scale that is above $\Lambda_{\rm QCD}$.
Degrees of freedom with masses much larger than this scale have been integrated out. When considering the amplitudes, the parameters are separated into three groups: well known ($G_F$, $\lambda_X$), perturbatively calculable ($c_i$), and non-perturbative ($\langle out |Q_i |in \rangle$).

In general, there is an infinite number of operators in the sum. When considering the quark transitions $q_h\to q_l$ ($b\to d$, $c\to u$, $s\to d$ for $P=B,D,K$, respectively), then, within the SM and to leading order in $G_F$, there are ten operators relevant to $P\to\pi\pi$ decays:
\begin{align}
&Q_1=(\bar q_{h\alpha}q_\beta)_{V-A}(\bar q_\beta q_{l\alpha})_{V-A},
&Q_2&=(\bar q_h q)_{V-A} (\bar qq_l)_{V-A},\no\\
&Q_3=(\bar q_hq_l)_{V-A}\mbox{$\sum_q$}(\bar qq)_{V-A},
&Q_4&=(\bar q_{h\alpha}q_{l\beta})_{V-A} \mbox{$\sum_q$}(\bar q_\beta q_\alpha)_{V-A},\no\\
&Q_5=(\bar q_hq_l)_{V-A}\mbox{$\sum_q$}(\bar qq)_{V+A},
&Q_6&=(\bar q_{h\alpha}q_{l\beta})_{V-A} \mbox{$\sum_q$}(\bar q_\beta q_\alpha)_{V+A}, \no\\
&Q_7=(3/2)\mbox{$\sum_q$} e_q(\bar q_hq_l)_{V-A}(\bar qq)_{V+A},
&Q_8&=(3/2)\mbox{$\sum_q$} e_q(\bar q_h q)_{V-A} (\bar qq_l)_{V+A},\no\\
&Q_9=(3/2)\mbox{$\sum_q$} e_q(\bar q_hq_l)_{V-A}(\bar qq)_{V-A},
&Q_{10}&=(3/2)\mbox{$\sum_q$} e_q(\bar q_hq)_{V-A} (\bar qq_l)_{V-A},
\end{align}
The operators are usually denoted as tree-level operators ($Q_{1,2}$), QCD penguin operators ($Q_{3,4,5,6}$), and electroweak penguin (EWP) operators ($Q_{7,8,9,10}$). 

%In what follows we discuss the 

%%%%%%%%%%%
%%%%%%%%%%%
%%%%%%%%%%%
\subsection{Truncating ${\cal H}_{\rm eff}$ with $Q_{1,\ldots,6}$}
\label{sec:truncate2}

As a first approximation for the $q_h\to q\bar q q_l$ transition amplitudes, we truncate ${\cal H}_{\rm eff}$ with the six leading operators, that is, the tree operators $Q_{1,2}$ and the strong penguin operators $Q_{3,4,5,6}$.
These six operators can be written in terms of $SU(2)_I$ representations~\cite{Bhattacharya:2025rrv}:
\beqa\label{eq:Q1234}
Q_1&=&-\frac{1}{\sqrt3}{\cal O}^{3/2}_{1/2}+\frac{1}{\sqrt6}{\cal O}^{1/2}_{1/2}+\frac{1}{\sqrt2}{\cal O}^{\prime1/2}_{1/2},\qquad
Q_2=-\frac{1}{\sqrt3}{\cal O}^{3/2}_{1/2}+\frac{1}{\sqrt6}{\cal O}^{1/2}_{1/2}-\frac{1}{\sqrt2}{\cal O}^{\prime1/2}_{1/2},\no\\
Q_3&=&\frac{\sqrt3}{\sqrt2}{\cal O}^{1/2}_{1/2}-\frac{1}{\sqrt2}{\cal O}^{\prime1/2}_{1/2}+{\cal O}^{\prime\prime1/2}_{1/2},\qquad\qquad
Q_4=\frac{\sqrt3}{\sqrt2}{\cal O}^{1/2}_{1/2}+\frac{1}{\sqrt2}{\cal O}^{\prime1/2}_{1/2}+{\cal O}^{\prime\prime1/2}_{1/2},
\eeqa
where the isospin basis operators correspond to ${\cal O}^{I}_{I_3}$.
The $Q_5$ and $Q_6$ operators have the same isospin decomposition as $Q_3$ and $Q_4$, respectively. For the $P^+\to\pi^+\pi^0$ decay, the initial state is 
$|1/2,+1/2 \rangle$ under $SU(2)_I$, and the final state is $\langle2,+1|$. Thus, only ${\cal O}^{3/2}_{1/2}$ can mediate this decay. (In principle, also ${\cal 
O}^{5/2}_{1/2}$ can do so, but it is not generated by the four-quark operators of the SM.) We learn that the QCD penguin operators $Q_{3,4,5,6}$ do not 
contribute to the charged meson decays.

In the absence of a strong penguin contribution, as is the case for $P^\pm\to\pi^\pm\pi^0$, there is only a single CKM factor, $\lambda_T$, at play, and thus
\beq\label{eq:acpzero}
A_{CP}(P^\pm\to\pi^\pm\pi^0)=0.
\eeq
We emphasize again that Eq.~(\ref{eq:acpzero}) is based on two approximations: First, truncating ${\cal H}_{\rm eff}$ with tree and strong penguin operators, and second, neglecting isospin violation in flavor conserving interactions. In the next section we go beyond the first approximation, and in Section~\ref{sec:isospinbreaking} beyond the second one. 
Since both effects are small, we treat them separately, that is, we assume isospin symmetry when we discuss the contributions of EWP operators, and we neglect EWP when we discuss other sources of isospin breaking.

%%%%%%%%%
%%%%%%%%%
%%%%%%%%%
\section{Electroweak penguins}
\label{sec:EWP}
In this section, we extend the operators we consider above and add to ${\cal H}_{\rm eff}$ the four electroweak penguin (EWP) operators relevant to $q_h\to q_l$ transitions, $Q_{7,8,9,10}$.
The expectation is that the inclusion of EWPs in the SM calculation leads to a direct CP asymmetry different from zero, but suppressed by the corresponding CKM factor and by ${\cal O}(\alpha)$. As we discuss below, the actual suppression turns out to be even stronger.

The EWP operators provide the ${\cal O}_{1/2}^{3/2}$ terms that contribute to the $P^+\to\pi^+\pi^0$ decay~\cite{Bhattacharya:2025rrv}:
\beq\label{eq:Q910}
Q_9=-\frac{\sqrt3}{2}{\cal O}^{3/2}_{1/2}-\frac{1}{\sqrt2}{\cal O}^{\prime1/2}_{1/2}-\frac{1}{2}{\cal O}^{\prime\prime1/2}_{1/2},\qquad
Q_{10}=-\frac{\sqrt3}{2}{\cal O}^{3/2}_{1/2}+\frac{1}{\sqrt2}{\cal O}^{\prime1/2}_{1/2}-\frac{1}{2}{\cal O}^{\prime\prime1/2}_{1/2}.
\eeq
The $Q_7$ and $Q_8$ operators have the same isospin decomposition as $Q_9$ and $Q_{10}$, respectively. The EWP terms also provide the required second CKM factor, $\lambda_P$, that interferes with the tree CKM factor $\lambda_{T}$.

When discussing $\Delta I=3/2$ transitions, Eqs. (\ref{eq:Q1234}) and (\ref{eq:Q910}) imply the following:
\begin{itemize}
\item The combination $(c_1-c_{2})(Q_1-Q_2)$ does not contribute,
    \item The strong penguin operators $Q_3,Q_4,Q_5,Q_6$ do not contribute,
    \item The combinations $(c_7-c_8)(Q_7-Q_8)$ and $(c_9-c_{10})(Q_9-Q_{10})$ do not contribute.
\end{itemize}
Thus, $(c_1+c_2)(Q_1^{(3/2)}+Q_2^{(3/2)})$, $(c_7+c_8)(Q_7^{(3/2)}+Q_8^{(3/2)})$ and $(c_9+c_{10})(Q_9^{(3/2)}+Q_{10}^{(3/2)})$ are left as the only three contributing combinations. 

The Wilson coefficients of the tree and EWP operators are provided in Table~\ref{tab:WCEWP}, based on Ref.~\cite{Buchalla:1995vs} (see also Ref.~\cite{Buras:1999st}).
%%%%%%%%%%%%%
\begin{table}[!t]
\begin{tabular}{|c|c|c|c|c|c|c|c|c|c|} \hline\hline
\rule{0pt}{1.2em}%
$P$ & $\mu$[GeV] & $c_1$ & $c_2$ & $c_7/\alpha$ & $c_8/\alpha$ & $c_9/\alpha$ & $c_{10}/\alpha$ & $\frac{c_9+c_{10}}{c_1+c_2}$ & $\frac{c_7+c_8}{c_9+c_{10}}$  \cr
\hline 
$B$ & $4.4$ & $-0.185$ & $+1.082$ & $-0.002$ & $+0.054$ & $-1.292$ & $+0.263$ & $-1.15\alpha$ & $-0.051$ \\
$D$ & $1.5$ & $-0.633$ & $+1.034$ & & & & & & \\
$K$ & $1.0$ & $-0.510$ & $+1.275$ & $-0.0032$ & $+0.173$ & $-1.576$ & $+0.690$ & $-1.16\alpha$ & $-0.19$ \\
\hline\hline
\end{tabular}
\caption{Wilson coefficients of tree and electroweak penguin operators (for $\Lambda_{\overline{\rm MS}}^{(5)}=225$ MeV, NDR). For $B$ and $K$, taken from Ref.~\cite{Buchalla:1995vs}, Tables VI, VII, XVIII and XXII. For $D$, taken from Ref.~\cite{deBoer:2016dcg}, Table 1.} 
\label{tab:WCEWP}
\end{table}   
%%%%%%%%%%%%%%%%%%
We learn that, within the SM,
\beq
\left|\frac{c_7+c_8}{c_9+c_{10}}\right|\ll1.
\eeq
Assuming that the hadronic matrix elements are of similar size, we first neglect the $Q_7+Q_8$ contribution.
Then, we can consider only the contributions from the tree, $Q_1+Q_2$, and the EWP, $Q_9+Q_{10}$, operators. 
But these two combinations fulfill
\beq
Q_1^{(3/2)}+Q_2^{(3/2)}=(3/2) \big(Q_9^{(3/2)}+Q_{10}^{(3/2)}\big).
\eeq
Considering the matrix element
\beq
\langle 2|{\cal H}_{\rm eff}^{3/2}|1/2\rangle=\lambda_T A_{3/2,2}^T+\lambda_P A_{3/2,2}^P,
\eeq
where ${\cal H}_{\rm eff}^{3/2}$ is the $\Delta I=3/2$ part, and
\beqa\label{eq:A32utfir}
A_{3/2,2}^T&=&-(1/\sqrt3)(c_1+c_2)\langle2|{\cal O}^{3/2}_{1/2}|1/2\rangle,\\
\label{eq:A32ut}
A_{3/2,2}^P&=&-(\sqrt3/2)(c_9+c_{10})\langle2|{\cal O}^{3/2}_{1/2}|1/2\rangle,
\eeqa
we obtain 
\beq
A_{3/2,2}^P=\frac32\frac{(c_9+c_{10})}{(c_1+c_2)}A_{3/2,2}^T.
\eeq
We learn that, neglecting $Q_{7,8}$, there is no strong phase between the EWP and tree contributions to $\Delta I=3/2$ transitions \cite{Neubert:1998pt,Bhattacharya:2025rrv}, so Eq.~(\ref{eq:acpzero}) still follows. (In Eq.~(\ref{eq:A32utfir}) we neglect the ${\cal O}(\alpha)$ contribution of $Q_9+Q_{10}$ to $A_{3/2,2}^T$. Including this contribution would not affect the strong phase alignment argument.) 

The different Lorentz structure of $Q_7+Q_8$ from that of $Q_9+Q_{10}$ implies that, in contrast to $Q_9+Q_{10}$, the $\Delta I=3/2$ matrix element of $Q_7+Q_8$ is not related to that of $Q_1+Q_2$, and in general differs by size and strong phase. Including $Q_7+Q_8$, Eq.~(\ref{eq:A32ut}) obtains an extra term:
\beq\label{eq:A32t78}
A_{3/2,2}^P=-(\sqrt3/2)(c_9+c_{10})\, \langle2|({\cal O}^{3/2}_{1/2})_{LL}|1/2\rangle\times(1+r_{78}e^{i\delta_{78}}),
\eeq
with
\beq\label{eq:r78del78}
r_{78}=\frac{c_7+c_8}{c_9+c_{10}}\left|\frac{\langle2|({\cal O}^{3/2}_{1/2})_{LR}|1/2\rangle}{\langle2|({\cal O}^{3/2}_{1/2})_{LL}|1/2\rangle}\right|,\qquad
\delta_{78}=\arg\left(\frac{\langle2|({\cal O}^{3/2}_{1/2})_{LR}|1/2\rangle}{\langle2|({\cal O}^{3/2}_{1/2})_{LL}|1/2\rangle}\right),
\eeq
where $r_{78}$ is real. Assuming that the hadronic matrix elements are of similar size, we have $r_{78}\ll1$. On the other hand, we do not have any information about $\delta_{78}$. The sine of the relative strong phase is then given by 
\begin{equation}
\sin\left[\arg\left(1+r_{78}e^{i\delta_{78}}\right)\right] = r_{78} \sin\delta_{78} + {\cal O}(r_{78}^2).
\end{equation}
The CP asymmetry  in $P^+\to\pi^+\pi^0$ to leading order in $r_{78}$ is now given by
\beq
A_{CP}(P^+\to\pi^+\pi^0)=3{\cal I}m\left(\frac{\lambda_P}{\lambda_T}\right)\frac{c_9+c_{10}}{c_1+c_2}\, r_{78}\sin\delta_{78}.
\eeq
Thus, three suppression factors are at play:
\begin{itemize}
    \item CKM suppression, ${\cal I}m\left(\lambda_P/\lambda_T\right)$.
    \item ${\cal O}(\alpha)$ suppression from $(c_9+c_{10})/(c_1+c_2)$.
    \item Strong phase suppression, $r_{78}$.
\end{itemize}
We note that the first two factors are reliably calculated. For the last one, the ratio of Wilson coefficients is known to be small. The overall suppression, however, relies on the assumption that the hadronic matrix elements are of comparable size.

%%%%%%%%%%
%%%%%%%%%%
%%%%%%%%%%
\section{Isospin breaking in QCD+QED}
\label{sec:isospinbreaking}
In flavor conserving interactions, there are two main isospin breaking parameters: The mass difference, $m_d-m_u$, and the EM charge difference, $Q_d-Q_u$. 
The size of the corresponding contributions are of order $(m_d-m_u)/\Lambda_{\rm QCD}\sim0.01$ and $\alpha\sim0.01$, respectively. Both accompany the operator $(u\bar u-d\bar d)$, and thus can be treated collectively as a spurion transforming as $\delta_{\rm IB}{\cal O}^1_0$ under $SU(2)_I$, with $\delta_{\rm IB}\sim0.01$.

There are three leading effects to the introduction of $\delta_{\rm IB}$:
\begin{itemize}
\item The $\pi^0$ mass eigenstate is no longer a pure $|1,0\rangle$ isospin eigenstate, as it acquires a $|0,0\rangle$ component (``$\pi$--$\eta$ mixing"). Consequently $\pi^+\pi^0$ acquires a $|1,1\rangle$ component.
\item The ${\cal O}^{1/2}_{1/2}{\cal O}^1_0$ operators have a $|3/2,1/2\rangle$ component, allowing the strong penguin operators to contribute to $P^+\to\pi^+\pi^0$ decays.
\item The ${\cal O}^{3/2}_{1/2}{\cal O}^1_0$ and ${\cal O}^{1/2}_{1/2}{\cal O}^1_0{\cal O}^1_0$ operators have a $|5/2,1/2\rangle$ component, allowing these operators to contribute to $P^+\to\pi^+\pi^0$ in this novel way.
\end{itemize}
Below we discuss these three effects in turn.

%%%%%%%%%
\subsection{$\pi$--$\eta$ mixing}

Isospin breaking results in $\pi^0$ mixing with $\eta$, $\eta'$ and other states. In the following, we consider only the $\pi$--$\eta$ mixing as it is the dominant effect.

The two relevant states of well-defined transformation properties under $SU(2)_I$ are $\pi_3(1,0)$, the EM-neutral member in the $SU(2)_I$-triplet, and $\pi_1(0,0)$, an $SU(2)_I$-singlet. We denote the rotation angle from the isospin basis $(\pi_3,\pi_1)$ to the mass basis $(\pi^0,\eta)$ by $\theta_{\pi\eta}$. To lowest order in chiral perturbation theory ($\chi$PT), it is given by \cite{Gross:1979ur}
\beq\label{eq:thetapieta}
\theta_{\pi\eta}=\frac{\sqrt3}{4}\frac{m_d-m_u}{m_s-\hat m}\simeq0.011,
\eeq
where $\hat m=(m_d+m_u)/2$. (Including higher order terms in $\chi$PT, there is a substantial uncertainty in this mixing angle; see, e.g., Refs.~\cite{Cirigliano:2003gt, Kroll:2005sd, Osipov:2025znd, FlavourLatticeAveragingGroupFLAG:2024oxs}.) Then,
\beq
\pi^0=c\pi_3+s\pi_0,\qquad \eta=c\pi_1-s\pi_3,
\eeq
where $c\equiv\cos\theta_{\pi\eta}$ and $s\equiv\sin\theta_{\pi\eta}$.

We define the decay amplitudes $A_3$ and $A_1$ as follows: 
\beq
A_3=A(P^+\to\pi^+\pi_3),\qquad A_1=A(P^+\to\pi^+\pi_1),
\eeq
so that
\beq
A(P^+\to\pi^+\pi^0)=cA_3+sA_1,\qquad A(P^+\to\pi^+\eta)=-sA_3+cA_1.
\eeq

The CP-averaged decay rates are given by
\beqa
\overline{\Gamma}(P^+\to\pi^+\pi^0)&=&\frac{|{\vec p}|}{8\pi m_P}\frac12\left[c^2(|A_3|^2+|\bar A_3|^2)
+2cs{\cal R}e(A_3 A_1^*+\bar A_3\bar A_1^*)+s^2(|A_1|^2+|\bar A_1|^2)\right],\no\\
\overline{\Gamma}(P^+\to\pi^+\eta)&=&\frac{|{\vec p}|}{8\pi m_P}\frac12\left[s^2(|A_3|^2+|\bar A_3|^2)
-2cs{\cal R}e(A_3 A_1^*+\bar A_3\bar A_1^*)+c^2(|A_1|^2+|\bar A_1|^2)\right].\qquad
\eeqa
Given the smallness of $\theta_{\pi\eta}$, Eq. (\ref{eq:thetapieta}), we take a zeroth order approximation:
\beq
\frac{|A_3|^2+|\bar A_3|^2}{|A_1|^2+|\bar A_1|^2}\simeq\frac{{\cal B}^{\pi^+\pi^0}}{{\cal B}^{\pi^+\eta}}.
\eeq
This estimate does not apply to $P=K$, where phase space does not allow $K\to\pi\eta$.

Considering the CP asymmetries, we have, to leading order in $s$, and taking into account that in the isospin limit and neglecting EWP operators $|A_3|=|\bar A_3|$, 
\beq
A_{CP}^{\pi^+\pi^0}=\frac{2s{\cal I}m(A_3A_1^*-\bar A_3\bar A_1^*)}{|A_3|^2+|\bar A_3|^2},\qquad
A_{CP}^{\pi^+\eta}=\frac{|A_1|^2-|\bar A_1|^2}{|A_1|^2+|\bar A_1|^2},
\eeq
and thus
\beq
\frac{A_{CP}^{\pi^+\pi^0}}{A_{CP}^{\pi^+\eta}}=\frac{2s{\cal I}m(A_3A_1^*-\bar A_3\bar A_1^*)}{|A_1|^2-|\bar A_1|^2}
\times\frac{|A_1|^2+|\bar A_1|^2}{|A_3|^2+|\bar A_3|^2}.
\eeq
The CKM factor ${\cal I}m(\lambda_T\lambda_P^*)$ cancels out in this ratio, and we obtain
\beq
\frac{A_{CP}^{\pi^+\pi^0}}{A_{CP}^{\pi^+\eta}}=2s\frac{{\cal I}m(A_3^T A_1^{P*}-A_3^P A_1^{T*})}{2{\cal I}m(A_1^T A_1^{P*})}\times
\frac{{\cal B}^{\pi^+\eta}}{{\cal B}^{\pi^+\pi^0}},
\eeq
where $A_i^{T,P}$ are the coefficients of $\lambda_{T,P}$.

Taking as a rough estimate, $|A_3^{T,P}/A_1^{T,P}|\sim({\cal B}^{\pi^+\pi^0}/{\cal B}^{\pi^+\eta})^{1/2}$, we obtain (again, a rough estimate)
\beq\label{eq:Apipithetapieta}
\frac{A_{CP}^{\pi^+\pi^0}}{A_{CP}^{\pi^+\eta}}\sim2s\, \frac{\sin\delta_{31}^{ut}-\sin\delta_{31}^{tu}}{2\sin\delta_{11}^{ut}}\times
\left(\frac{{\cal B}^{\pi^+\eta}}{{\cal B}^{\pi^+\pi^0}}\right)^{1/2},
\eeq
where $\delta_{ij}^{qp}$ are strong phases. Assuming that the ratio involving strong phases is ${\cal O}(1)$, we thus estimate, for $P=B,D$,
\beq\label{eq:acpeta}
\left|A_{CP}(P^\pm\to\pi^\pm\pi^0)\right| \sim0.02\, ({\cal B}^{\pi^+\eta}/{\cal B}^{\pi^+\pi^0})^{1/2} \left|A_{CP}(P^\pm\to\pi^\pm\eta)\right|.
\eeq

The current experimental ranges for the branching fractions and CP asymmetries of the $\pi^+\eta$ modes are presented in Table~\ref{tab:pietaexp}. The PDG average for $B$ decay is based on measurements by Belle \cite{Belle:2011hjx} and BaBar \cite{BaBar:2009cun}, and for $D$ decay on measurements by LHCb \cite{LHCb:2022pxf}, Belle \cite{Belle:2011tmj}, and CLEO \cite{CLEO:2009fiz}.
%%%%%%%%%%%%%
\begin{table}[!t] 
\begin{tabular}{|c|c|c|} \hline\hline
\rule{0pt}{1.2em}%
$P$ & ${\cal B}(P^+\to\pi^+\eta)$ & $A_{CP}$  \cr
\hline 
$B$ & $(4.02\pm0.27)\times10^{-6}$ & $-0.143\pm0.051$ \\
$D$ & $(3.77\pm0.09)\times10^{-3}$ & $-0.003\pm0.005$ \\
\hline\hline
\end{tabular}
\caption{Experimental values of CP asymmetries and CP-averaged branching fractions $P^+\to\pi^+\eta$ for $P=B$~\cite{HeavyFlavorAveragingGroupHFLAV:2024ctg} and $D$ \cite{ParticleDataGroup:2024cfk}.}
	\label{tab:pietaexp}
\end{table}   
%%%%%%%%%%%%%%%%%%

%%%%%%%%%
\subsection{${\cal O}^{3/2}_{1/2}$ from $Q_{3,4,5,6}$}
The product of an operator ${\cal O}^{1/2}_{1/2}$ with a spurion $\delta_{\rm IB}{\cal O}^1_0$ generates an effective operator ${\cal O}^{3/2}_{1/2}$ suppressed by $\delta_{\rm IB}\sim0.01$. Coming from the $Q_{3,4,5,6}$ operators, it gives a contribution accompanied by the CKM factor $\lambda_P$. Then, it contributes to the $P^+\to\pi^+\pi^0$ decay via
\beq\label{eq:apsp}
A^P_{3/2,1/2}\sim \delta_{\rm IB}c_i\langle 2|({\cal O}_i)^{1/2}_{1/2}{\cal O}^1_0|1/2\rangle,\ \ \ i=3,4,5,6.
\eeq
The Wilson coefficients of the strong penguin operators are presented in Table~\ref{tab:WCSP}.
%%%%%%%%%%%%%
\begin{table}[!t]
%	\begin{center}
		\begin{tabular}{|c|c|c|c|c|c|} \hline\hline
			\rule{0pt}{1.2em}%
			$P$ & $\mu$ [GeV] & $c_3$ & $c_4$ & $c_5$ & $c_6$ \cr
			\hline 
			$B$ & $4.4$ & $+0.014$ & $-0.035$ & $+0.009$ & $-0.041$ \\
            $D$ & $1.3$ & $-0.008$ & $-0.093$ & $+0.000$ & $+0.001$ \\
            $K$ & $1.0$ & $+0.032$ & $-0.058$ & $-0.001$ & $-0.111$ \\
			\hline\hline
		\end{tabular}
%	\end{center}
\caption{Wilson coefficients of QCD penguin operators. For $B$ and $K$ decays, taken from Ref.~\cite{Buchalla:1995vs}, Tables XVIII and XXII. For $D$ decays, taken from Ref.~\cite{deBoer:2016dcg}, Table 1.} 
	\label{tab:WCSP}
\end{table}   
%%%%%%%%%%%%%%%%%%

The same Wilson coefficients play a role in the neutral $P^0\to\pi\pi$ decays. Thus, if all hadronic matrix elements were of similar size, the strong penguin contributions of Eq.~(\ref{eq:apsp}) to $A_{CP}(P^+\to\pi^+\pi^0)$ would be suppressed, compared to the direct CP asymmetries $A_{CP}(P^0\to\pi\pi)$:
\beq
A_{CP}(P^+\to\pi^+\pi^0)\sim\delta_{\rm IB}A_{CP}(P^0\to\pi\pi).
\eeq
The current values of the direct CP asymmetries in neutral meson decays are presented in Table~\ref{tab:acpneutral}. The PDG average for $B^0\to\pi^+\pi^-$ is based on measurements by LHCb~\cite{LHCb:2020byh,LHCb:2018pff}, Belle~\cite{Belle:2013epq} and BaBar~\cite{BaBar:2012fgk}, for $B^0\to\pi^0\pi^0$ on measurements by Belle II~\cite{Belle-II:2023cbc}, Belle~\cite{Belle:2017lyb}, and BaBar~\cite{BaBar:2012fgk} (for a more recent Belle II result, see \cite{Belle-II:2024baw}). 
The asymmetry in $D^0\to\pi^+\pi^-$ is based on an LHCb measurement~\cite{LHCb:2022lry} and in $D^0\to\pi^0\pi^0$ on a Belle measurement~\cite{Belle:2014evd} (for a more recent Belle~II measurement, see \cite{Belle-II:2025rmf}).
%%%%%%%%%%%%%
\begin{table}[!t]
%	\begin{center}
		\begin{tabular}{|c|c|c|} \hline\hline
			\rule{0pt}{1.2em}%
			$P$ & $\pi^+\pi^-$ & $\pi^0\pi^0$  \cr
			\hline 
			$B$ & $+0.314\pm0.030$ & $+0.25\pm0.20$ \\
            $D$ & $(-2.3\pm0.6)\times10^{-3}$ & $(+0.3\pm6.4)\times10^{-3}$ \\
            $K$ & $(-5.2\pm0.8)\times10^{-6}$ & $(+1.04\pm0.16)\times10^{-5}$ \\
			\hline\hline
		\end{tabular}
%	\end{center}
	\caption{Measurements of CP asymmetries in $P^0\to\pi^+\pi^-$ and $P^0\to\pi^0\pi^0$ for $P=B$ \cite{ParticleDataGroup:2024cfk} (note that the sign convention for $C_{\pi\pi}$ in Ref.~\cite{ParticleDataGroup:2024cfk} is opposite to our sign convention for $A_{CP}$); $P=D$ \cite{HeavyFlavorAveragingGroupHFLAV:2024ctg}; and $P=K$ (see Appendix~\ref{app:epsp}).} 
	\label{tab:acpneutral}
\end{table}   
%%%%%%%%%%%%%%%%%%

%%%%%%%%%
\subsection{${\cal O}^{5/2}_{1/2}$ from $Q_{1,2}$ and from $Q_{3,4,5,6}$}
The possibility of generating ${\cal O}^{5/2}_{1/2}$ operators from tree operators, suppressed by $\delta_{\rm IB}$, would provide only a small correction to our estimates. The reason is that this contribution is accompanied by the CKM factor $\lambda_T$, to which $A_{3/2,1/2}^T$ contributes with no such suppression. 

The possibility of generating ${\cal O}^{5/2}_{1/2}$ operators from QCD penguin operators, suppressed by $(\delta_{\rm IB})^2$, would only provide a small correction to our estimates. While it provides a contribution accompanied by the CKM factor $\lambda_P$, it is $\delta_{\rm IB}$-suppressed compared to the contribution of Eq.~(\ref{eq:apsp}).

Thus, either contribution is only a correction of ${\cal O}(\delta_{\rm IB})$ to other contributions and, therefore, negligible. (See discussions in Ref.~\cite{Botella:2006zi} for $B\to\pi\pi$, and in Ref.~\cite{Gardner:2000sb} in the context of $K\to\pi\pi$.)

%%%%%%%%%%
%%%%%%%%%%
%%%%%%%%%%
\section{The CP asymmetries in charged $B$, $D$ and $K$ decays}
\label{sec:BDK}
In this section, we apply the formalism presented above to the concrete cases of $B^+$, $D^+$, and $K^+$ decays into $\pi^+\pi^0$, and provide a quantitative estimate for each.

%%%%%%%%%%%
\subsection{$A_{CP}(B^\pm\to\pi^\pm\pi^0)$}
\label{sec:B}

The CP asymmetry in $B\to\pi\pi$ is not CKM suppressed:
\beq
{\cal I}m\left(\frac{\lambda_P}{\lambda_T}\right) = 
{\cal I}m\left(\frac{V_{tb}^*V_{td}}{V_{ub}^*V_{ud}}\right)=\sin\alpha\approx0.99.
\eeq

The contribution of the EWP operators is suppressed not only by the expected ${\cal O}(\alpha)$ factor in $(c_9+c_{10})/(c_1+c_2)$, but also by the strong phase alignment which introduces an additional suppression factor, $(c_7+c_8)/(c_9+c_{10})$. Together, the suppression factor is given by (see Table~\ref{tab:WCEWP})
\beqa\label{eq:ewpb}
A_{CP}(B^\pm\to\pi^\pm\pi^0)&\sim&\frac{3(c_7+c_8)}{c_1+c_2}\no\\
&\approx& 1.3\times10^{-3}.
\eeqa

The contribution generated by $\pi$--$\eta$ mixing \cite{Gronau:2005pq,Gardner:2005pq} can be estimated from Eq.~(\ref{eq:acpeta}) and the data in Table~\ref{tab:pietaexp}:
\beqa\label{eq:pietab}
A_{CP}(B^\pm\to\pi^\pm\pi^0)&\sim&0.02 \left[{\cal B}(B^+\to\pi^+\eta)/{\cal B}(B^+\to\pi^+\pi^0)\right]^{1/2}A_{CP}(B^\pm\to\pi^\pm\eta)\no\\*
&\approx&(2.5\pm0.9)\times10^{-3},
\eeqa
currently providing a $2.8\sigma$ hint of a nonzero contribution.

The contribution generated by isospin breaking in the strong penguin operators can be estimated by the CP asymmetry in neutral $B$ decays:
\beqa\label{eq:IBPb}
A_{CP}(B^\pm\to\pi^\pm\pi^0)&\sim&\delta_{\rm IB}\times A_{CP}(B^0\to\pi^+\pi^-)\no\\
&\approx&3\times10^{-3},
\eeqa
where we used the experimental value from Table \ref{tab:acpneutral}.
To summarize, Eq.~(\ref{eq:ewpb}) tells us that EWP operators contribute at a level of $10^{-3}$, Eq.~(\ref{eq:pietab}) tells us that the contribution of $\pi$--$\eta$ mixing is $\sim {\rm few} \times 10^{-3}$ (depending on the not yet established $A_{CP}(B^\pm\to\pi^\pm\eta)$), and Eq.~(\ref{eq:IBPb}) tells us that isospin violation in the QCD penguin contributions yields ${\cal O}({\rm few} \times 10^{-3})$ CP asymmetry.

We conclude that the CP asymmetry $A_{CP}(B^\pm\to\pi^\pm\pi^0)$ is generated by three different effects, of order $\alpha\alpha_s$, $\theta_{\pi\eta}\alpha_s$, and $r_{78}\alpha$, which all happen to be of a similar order of magnitude. It is expected then to be of ${\cal O}(0.003)$, with large uncertainties from hadronic matrix elements and strong phases. 

While there are many statements in the literature that the asymmetry is small, only a few works provide explicit numerical estimates. Ref.~\cite{Beneke:2003zv} uses QCD factorization to estimate the electroweak penguin (EWP) contribution and quotes a central value of $-2 \times 10^{-4}$, with theoretical uncertainties of order $10^{-3}$. Ref.~\cite{Bauer:2005kd} employs SCET and finds that the asymmetry is below $0.05$. Ref.~\cite{Huber:2021cgk} states that the asymmetry is very small, and their global fit yields $0.05\pm0.20$. 
A recent global fit yields $-0.024 \pm 0.016$~\cite{Fang:2026fhl}, a factor of a few above our estimate. 
The difference is due to the value of $r_{78}$. In the notation of our paper, the fit in Ref.~\cite{Fang:2026fhl} gives a value of $r_{78}$ that is much larger than the one we used. As stated below Eq.~(\ref{eq:r78del78}), our estimate is based on the assumption that the two relevant hadronic matrix elements are comparable, while the central value from the fit of Ref.~\cite{Fang:2026fhl} implies that the ratio of hadronic matrix elements is much larger than 1.

%%%%%%%%%%
%%%%%%%%%%
%%%%%%%%%%
\subsection{$A_{CP}(D^\pm\to\pi^\pm\pi^0)$}
\label{sec:D}
The CP asymmetry in $D\to\pi\pi$ is strongly CKM suppressed:
\beq\label{eq:ckmD}
{\cal I}m\left(\frac{\lambda_P}{\lambda_T}\right)={\cal I}m\left(\frac{V_{cb}^*V_{ub}}{V_{cd}^*V_{ud}}\right)\approx 7.0\times10^{-4}\sin\gamma\approx6.4\times10^{-4}.
\eeq

Other suppression factors are challenging to estimate because long distance contributions to $D\to\pi\pi$ are significant and can even be dominant (see, {\it e.g.}, Refs.~\cite{Golden:1989qx,Brod:2011re}). The short distance contributions of the penguin operators are suppressed not only by ${\cal O}(\alpha)\sim0.01$, but also by the GIM mechanism, roughly a factor of ${\cal O}(m_b^2/m_W^2)\sim0.003$. 

The estimation of the EWP contribution to the CP asymmetry is particularly difficult. The naive estimate combines the CKM suppression of Eq. (\ref{eq:ckmD}) with the $\alpha$-suppression of EWP-to-tree ratio, suggesting that this contribution to the CP asymmetry is $\sim{\rm few}\times10^{-6}$. 

The contribution generated by $\pi$--$\eta$ mixing can be estimated from Eq.~(\ref{eq:acpeta}) and the data in Table~\ref{tab:pietaexp}:
\beqa\label{eq:pietad}
A_{CP}(D^\pm\to\pi^\pm\pi^0)&\sim&0.02 \left[{\cal B}(D^+\to\pi^+\eta)/{\cal B}(D^+\to\pi^+\pi^0)\right]^{1/2}A_{CP}(D^\pm\to\pi^\pm\eta)\no\\*
&=&(1.0\pm1.6)\times10^{-4},
\eeqa
currently providing only an upper bound on this contribution. The two relevant suppression factors -- the CKM suppression of Eq.~(\ref{eq:ckmD}) and the $\pi$--$\eta$ mixing of Eq.~(\ref{eq:thetapieta}) -- suggest that also this contribution is $\sim{\rm few}\times10^{-6}$.

The contribution generated by isospin breaking in the strong penguin operators can be estimated by the CP asymmetry in neutral $D$ decays:
\beq\label{eq:IBPd}
A_{CP}(D^\pm\to\pi^\pm\pi^0)\sim\delta_{\rm IB}\times A_{CP}(D^0\to\pi^+\pi^-)
\approx2\times10^{-5},
\eeq
where we used the experimental value from Table \ref{tab:acpneutral}. Note that the CP asymmetry in $D^0\to\pi^+\pi^-$ is a factor of three larger than even just the CKM suppression factor of Eq.~(\ref{eq:ckmD}). There has been much discussion in the literature whether this surprisingly large asymmetry is due to new physics or to a QCD effect. 
Either way, it implies a strong enhancement, $|A^P/A^T|\gg1$ (for an explicit estimate, see Ref.~\cite{Gavrilova:2023fzy}), which we expect to affect also the asymmetry in the charged $D$ decay.

We conclude that the CP asymmetry $A_{CP}(D^\pm\to\pi^\pm\pi^0)$ is likely to be dominated by isospin violation in the strong interactions. It is expected to be of ${\cal O}(10^{-5})$, with large uncertainties from hadronic matrix elements and strong phases. 

Previous studies \cite{Grossman:2006jg,Brod:2011re,Cheng:2012wr,Grossman:2012eb} did not go beyond ``the isospin limit" prediction of vanishing asymmetry.

%%%%%%%%%%
%%%%%%%%%%
%%%%%%%%%%
\subsection{$A_{CP}(K^\pm\to\pi^\pm\pi^0)$}
\label{sec:K}
The CP asymmetry in $K\to\pi\pi$ is strongly CKM suppressed:
\beq\label{eq:ckmK}
{\cal I}m\left(\frac{\lambda_P}{\lambda_T}\right)={\cal I}m\left(\frac{V_{ts}^*V_{td}}{V_{us}^*V_{ud}}\right)=1.6\times10^{-3}\sin\beta\approx6.2\times10^{-4}.
\eeq

The contribution of EWP operators is suppressed not only by ${\cal O}(\alpha)$ (see the $(c_9+c_{10})/(c_1+c_2)$ entry in Table~\ref{tab:WCEWP}), but also by a tiny strong phase. In fact, this suppression is much stronger than in the corresponding $B$-meson decay. It follows from the Watson theorem \cite{Watson:1952ji}: If $\pi-\pi$ scattering is elastic, then the strong phase in $K\to\pi\pi$ decays depends only on the isospin of the final state. Indeed, the inelasticity in $\pi-\pi$ scattering is known to be small at $E\sim m_K$ \cite{Caprini:2005zr,Garcia-Martin:2011iqs,Hyams:1973zf,Pennington:1973xv}, $\lesssim10^{-4}$ for final $I=2$ states due to the absence of corresponding resonances.  We conclude that the contribution of EWP operators to the CP asymmetry in $K^\pm\to\pi^\pm\pi^0$ is $\lesssim10^{-9}$.

The contribution from $\pi$--$\eta$ mixing cannot be estimated in a way 
similar to that for the $B$ and $D$ decays, because, due to phase space, there is no $K\to\pi\eta$ decay. We can only make an estimate on the basis of the known suppression factors: The CKM suppression of Eq.~(\ref{eq:ckmK}) and the $\pi$--$\eta$ mixing angle of
Eq.~(\ref{eq:thetapieta}). Their combination implies that the contribution of the $I=1$ component in the $\pi^+\pi^0$ state is $\lesssim6\times10^{-6}$.

The contribution generated by isospin breaking in the strong penguin operators can be estimated by the CP asymmetry in neutral $K$ decays:
\beq\label{eq:IBPk}
A_{CP}(K^\pm\to\pi^\pm\pi^0)\sim\delta_{\rm IB}\times A_{CP}(K^0\to\pi\pi)
\approx2\times10^{-7},
\eeq
where we used the experimental value from Table \ref{tab:acpneutral}. 

We conclude that $A_{CP}(K^\pm\to\pi^\pm\pi^0)$ gets a negligible contribution from EWP operators, and is dominated by isospin violation in QCD, particularly by the contribution related to the $I=1$ component in $\pi^+\pi^0$ stemming from $\pi$--$\eta$ mixing. It is expected to be of ${\cal O}(10^{-5}-10^{-6})$, with large uncertainty coming from the strong phase of the $I=1$ final state.

A similar conclusion is stated in Refs.~\cite{Dib:1990gr,Dib:1990qj}, where an estimate of ${\cal O}(10^{-6})$ is quoted with large uncertainties. In particular, they argue that electromagnetic corrections to the weak amplitude give the dominant contribution, leading to an asymmetry at the level of $10^{-6}$, while isospin-breaking effects from quark mass differences do not contribute. While we agree with their numerical estimate, we differ in the identification of the leading effect.
Ref.~\cite{Riazuddin:1993pn} studies the effect of intermediate vector meson exchange in generating the asymmetry. Using updated values of the CKM factors, we find that their estimate corresponds to an upper bound of order $10^{-5}$.

%%%%%%%%%%
%%%%%%%%%%
%%%%%%%%%%
\section{Conclusions}
\label{sec:conclusions}
There are three possible sources of CP asymmetries in charged meson decays into two pions. Electroweak penguins (EWP) and isospin breaking in QCD penguins  are both suppressed, but are roughly of the same order. The $\Delta I=5/2$ contribution is suppressed by second-order isospin breaking or by second order in the weak interaction, and is much smaller than the other two.

In our analysis, some of our arguments are based on experimental data and spurion analysis. This is the case in estimating the contribution from $\pi$--$\eta$ mixing using the experimental value of $A_{CP}(P^\pm\to\pi^\pm\eta)$ and in estimating the contribution from isospin breaking in strong penguins using the experimental value of $A_{CP}(P^0\to\pi^+\pi^-)$. These estimates are then valid also beyond the Standard Model. In cases where we use explicit SM input, such as the estimate of EWP operators using the SM values of the relevant Wilson coefficients, new physics can modify these predictions.

The order of magnitude estimates of the yet-unmeasured CP asymmetries $A_{CP}(P^\pm\to\pi^\pm\pi^0)$ are presented in Table~\ref{tab:acpBDK}.
We estimate the size of the asymmetries to be
\begin{align}
A_{CP}(B^+\to\pi^+\pi^0)&\sim3\times10^{-3}, \nonumber \\
A_{CP}(D^+\to\pi^+\pi^0)&\sim10^{-5}, \nonumber \\
A_{CP}(K^+\to\pi^+\pi^0)&\sim10^{-6}.
\end{align}
For the CP asymmetries in $D$ and $K$ decays, there are large hadronic uncertainties, and each could be an order of magnitude different (in either direction -- enhancement or suppression) from these estimates. For the CP asymmetry in $B$ decay there are several contributions of order a few permille which can yield a total asymmetry in the range $0.001-0.01$. 

Let us highlight some of the unique features for each of the decaying mesons:
\begin{itemize} 
\item $A_{CP}(B^\pm\to\pi^\pm\pi^0)$: Perturbative calculations are possible. They imply, in particular, that there is an approximate strong phase alignment between tree and EWP operators, which suppresses the EWP contribution to the CP asymmetry well below the naive estimate. Measuring $A_{CP}(B^\pm\to \pi^\pm\eta)$ would allow a better estimate of the contribution from $\pi$--$\eta$ mixing.
\item $A_{CP}(D^\pm\to\pi^\pm\pi^0)$: The CKM structure and the quark mass hierarchy combine to imply that long distance physics plays a significant role, and makes an estimate of the asymmetry particularly challenging. The experimental measurement of $A_{CP}(D^0\to\pi^+\pi^-)$ combined with a spurion analysis of isospin breaking provide the best handle over the expected size of the asymmetry. Measuring $A_{CP}(D^\pm\to\pi^\pm\eta)$ would allow a better estimate of the contribution from $\pi$--$\eta$ mixing.
\item $A_{CP}(K^\pm\to\pi^\pm\pi^0)$: The Watson theorem implies that all contributions to the final $I=2$ state carry the same strong phase, and thus do not generate an asymmetry. Consequently, $\pi$--$\eta$ mixing plays a crucial role, as it adds a final $I=1$ component, accompanied by an independent strong phase.
\end{itemize}

These considerations warrant increased experimental efforts to measure $A_{CP}(P^\pm\to\pi^\pm\pi^0)$ for $P=B,D,K$, as there is much to learn from improved bounds or measurements in the future. In particular,
\begin{itemize}
    \item Measuring $A_{CP}(B^\pm\to\pi^\pm\pi^0)$ has direct implications for the extraction of $\alpha$ from $B\to\pi\pi$ decays~\cite{Gronau:1990ka,Gardner:2005pq}, and indirect implications for the so-called $K\pi$ puzzle that is also related to isospin violation~\cite{Gronau:2005kz}.
    \item Measuring $A_{CP}(D^\pm\to\pi^\pm\pi^0)$ may shed light on the controversy regarding the surprisingly large value of $\Delta A_{CP}$ (see, {\it e.g.}, Refs.~\cite{Grossman:2019xcj,Dery:2019ysp,Schacht:2022kuj,Sinha:2025cuo,Friday:2025gpj,Fleischer:2025zhl}).
    \item Measuring $A_{CP}(K^\pm\to\pi^\pm\pi^0)$ will provide a unique experimental probe of $\pi$--$\eta$ mixing.
\end{itemize}

%%%%%%%%%%%%%
\begin{table}[!t]
%	\begin{center}
		\begin{tabular}{|c|c|c|c|c|} \hline\hline
			\rule{0pt}{1.2em}%
			$P$ & EWP & $\theta_{\pi\eta}$ & ${\cal O}^{1/2}_{1/2}{\cal O}^1_0$ & $A_{CP}$  \cr
			\hline 
			$B$ & $10^{-3}$ & $10^{-3}$ & $10^{-3}$ & $10^{-3}$ \\
            $D$ & $10^{-6}$ & $10^{-6}$ & $10^{-5}$ & $10^{-5}$ \\
            $K$ & $10^{-9}$ & $10^{-6}$ & $10^{-7}$ & $10^{-6}$ \\
			\hline\hline
		\end{tabular}
%	\end{center}
	\caption{Order of magnitude estimates of the contribution of various mechanisms to $A_{CP}(P^\pm\to\pi^\pm\pi^0)$, for $P=B,D,K$: EWP operators, $\pi$--$\eta$ mixing, isospin violation in strong interactions, and our estimate of the total SM prediction.} 
	\label{tab:acpBDK}
\end{table}   
%%%%%%%%%%%%%%%%%%

%%%%%%%%%%%%%%%%
\section*{Acknowledgements}
We thank Gerhard Buchalla, Margarita Gavrilova, Stefan Schacht, and Antonio Pich for useful discussions. 
YG is supported by the NSF grant PHY-2309456. 
ZL is supported in part by the Office of High Energy Physics of the U.S.\ Department of Energy under contract DE-AC02-05CH11231. 
YN is supported by a grant from the Minerva Foundation (with funding from the Federal Ministry for Education and Research).

%%%%%%%%%%%%%%%%%%%%%%%%
\appendix
\section{CP violation in decay for $K\to\pi^0\pi^0$ and $K\to\pi^+\pi^-$}
\label{app:epsp}
CP violation in decay in neutral $K\to\pi\pi$ decays is given by ${\cal R}e(\epsilon^\prime)$ \cite{ParticleDataGroup:2024cfk}, where $\epsilon^\prime$ is defined via
\beqa
\eta_{00}&=&\frac{1-\lambda_{00}}{1+\lambda_{00}}=\epsilon-2\epsilon^\prime,\no\\
\eta_{+-}&=&\frac{1-\lambda_{+-}}{1+\lambda_{+-}}=\epsilon+\epsilon^\prime,
\eeqa
and
\beq
\lambda_f=(q/p)(\overline{A}_f/A_f),
\eeq
and $q/p$ are $K^0-\overline{K}{}^0$ mixing parameters. We obtain:
\beq
\epsilon^\prime\simeq\frac16(\lambda_{00}-\lambda_{+-}).
\eeq
Using
\beqa\label{eq:A00Apm}
A_{\pi^0\pi^0}=\sqrt{\frac13}|A_0|e^{i(\delta_0+\phi_0)}
-\sqrt{\frac23}|A_2|e^{i(\delta_2+\phi_2)},\no\\
A_{\pi^+\pi^-}=\sqrt{\frac23}|A_0|e^{i(\delta_0+\phi_0)}
+\sqrt{\frac13}|A_2|e^{i(\delta_2+\phi_2)},
\eeqa
with $\delta_i$ strong and $\phi_i$ weak phases, we obtain:
\beq
{\cal R}e(\epsilon^\prime)=-\frac{1}{\sqrt{2}}\left|\frac{A_2}{A_0}\right|\sin(\delta_2-\delta_0)\sin(\phi_2-\phi_0).
\eeq
We thus identify $A_{CP}(K\to\pi^+\pi^-)=-2{\cal R}e(\epsilon^\prime)$ and $A_{CP}(K\to\pi^0\pi^0)=+4{\cal R}e(\epsilon^\prime)$. 

The ratio of $-2$ between the CP asymmetries in the two modes is a result of the CG coefficients in Eq.~(\ref{eq:A00Apm}) and the expansion of the CP asymmetries to first order in $|A_2/A_0|$. It does not hold in the corresponding $B$ and $D$ decays, because there $|A_2/A_0|\not\ll1$.

The measured observable is ${\cal R}e(\epsilon^\prime/\epsilon)$. To extract ${\cal R}e(\epsilon^\prime)$ we notice the following:
\begin{enumerate}
\item Accidentally, $\arg(\epsilon)\approx\arg(\epsilon^\prime)$ and, consequently, ${\cal R}e(\epsilon^\prime/\epsilon)\simeq|\epsilon^\prime/\epsilon|$.
\item Experimentally, $\delta_2-\delta_0\approx\pi/4$ and, consequently, ${\cal R}e(\epsilon^\prime)/|\epsilon^\prime|\approx\cos(\pi/4)$.
\end{enumerate}
We obtain:
\beq
{\cal R}e(\epsilon^\prime)\simeq{\cal R}e(\epsilon^\prime/\epsilon)\times|\epsilon|\times\cos(\pi/4).
\eeq
%

%%%%%%%%%%%%%%%%%%%%%%%%

\end{document}